# Identifying Highly Deformable van der Waals Layered Chalcogenides with Superior Thermoelectric Performance Using Deformability Factors and Interpretable Machine Learning

Qi Ren[a], Yingzhuo Lun[a], Bonan Zhu[a], Gang Tang[b,*], and Jiawang Hong[a,*]

[a] School of Aerospace Engineering, Beijing Institute of Technology, Beijing, 100081, China

[b] Advanced Research Institute of Multidisciplinary Science, Beijing Institute of Technology, Beijing 100081, China.

[*] Corresponding author E-mail:

gtang@bit.edu.cn (Gang Tang),

hongjw@bit.edu.cn (Jiawang Hong)

**ABSTRACT**

 Van der Waals layered chalcogenide-based flexible thermoelectric devices show great potential for applications in wearable electronics. However, materials that are both highly deformable and exhibit superior thermoelectric performance are extremely limited. There is an urgent need for methods that can efficiently predict both deformability and thermoelectric performance to enable high-throughput screening of these materials. In this study, over 1000 van der Waals layered chalcogenides were high-throughput screened from material databases, the deformability of which were predicted with our previously developed deformability factor. An accurate and efficient model based on machine learning methods were developed to predict the thermoelectric properties. Several candidate materials with both deformability and thermoelectric potential were successfully discovered. Among them, $NbSe_2Br_2$ was verified by first principles calculations, achieving $ZT_{max}$ value of 1.35 at 1000K, which is currently the highest value among flexible inorganic thermoelectric materials. And the power factor value of 8.1 $\mu Wcm^{-1}K^{-2}$ at 300K also surpassed most organic and inorganic flexible thermoelectric materials. Its high deformability mainly attributed to the small slipping energy that allows interlayer slip and the small in-plane modulus that allows deformation before failure. The high $ZT_{max}$ is mainly contributed by the extremely low thermal conductivity and the high Seebeck coefficient along the out-of-plane direction at high temperature. The high power factor at room temperature is mainly comes from the high conductivity in the in-plane direction. This study is expected to accelerate the development and application of flexible thermoelectric devices based on inorganic semiconductor materials.

## INTRODUCTION

Wearable electronic devices are widely applied in fields such as communication, health monitoring, robotics, and other areas[1]. However, their utility is significantly constrained due to the limited energy capacity and short lifetime of chemical batteries[2]. It's urgently desired to find more portable, reliable, super-thin, and sustainable flexible power supplies[3,4]. Flexible thermoelectric (TE) generators can directly convert body heat into electricity, providing an effective self-power supply solution[5]. Flexible TE technology requires high performance flexible TE materials to fabricate high power density flexible TE device[6]. Semiconducting organic polymers, due to their inherent softness and flexibility, are candidate materials for flexible TE materials, but they generally have poor electrical transport properties and TE performance[7–9]. In contrast, inorganic TE materials have higher TE performance, but they are inherent brittleness to withstand deformation or mechanical processing[10,11]. Therefore, there is an urgent need to develop new materials with both high deformability and superior TE performance.

Recently, the discovery of inorganic ductile TE semiconductors based on van der Waals (vdW) layered chalcogenides open a new avenue to fabricate high performance flexible TE materials. For example, inorganic $Ag_2S$–based[12] materials and vdW layerd chalcogenide single-crystalline InSe[13] with good ductility and tunable electrical properties have been reported. Based on this development, AgCu(Se,S,Te) pseudoternary solid solutions[2], with the highest figure-of-merit (ZT) values in flexible TE materials (0.45 at 300 K and 0.68 at 340 K) were developed, further demonstrated thin and flexible p-shaped devices with a maximum normalized power density that reaches 30 μW cm$^{−2}$ K$^{−2}$. VdW layered chalcogenide single-crystalline $SnSe_2$[6] also exhibits good plasticity as well as a high power factor (PF) of 10.8 μW cm$^{−1}$ K$^{−2}$ at room temperature and a ZT of 0.09 at 375 K along the in-plane direction upon doping a tiny amount of halogen elements. The room-temperature PF is about two times of the maximum PF values of the flexible organic TE materials and flexible inorganic $Ag_2S$-based TE materials reported before.

On the one hand, a variety of vdW layered chalcogenides exhibit high TE performance, including SnSe[14,15] with the recorded highest ZT value of 2.6 at 923 K, the widely commercialized room-temperature TE material $Bi_2Te_3$[16], GeTe[17], BiCuOSe[18] and so on. On the other hand, the exceptional plasticity discovered in InSe crystal indicates the existence of abundant plastically deformable vdW layered chalcogenides. In order to high-throughput screen candidate materials, a deformability factor which can quantify the plastic deformability of materials was proposed[13]. A nearly automated and efficient high-throughput screening methodology was used and tens of potential 2D chalcogenide crystals with plastic deformability were discovered[19]. In our previous work, a simple and improved plastic deformability factor was put forward, with which 99 types of vdW layered materials with high deformability factors were screened out from more than 40 000 materials[20]. These achievements inspire the great enthusiasm in finding new ductile TE materials based on vdW layered chalcogenides toward flexible TE devices and applications. However, the variety of candidates discovered to date is very limited, the conventional trial-and-error method is too time-consuming and costly. The lack of an efficient method to predict TE properties also greatly restricts the development efficiency of flexible TE devices.

In this work, over 1000 vdW layered chalcogenides were screened from material databases by high-throughput screening methodology, the deformability of which were predicted with our previously developed deformability factor[20]. An accurate and efficient model based on machine learning (ML) methods were developed to predict the TE properties. Several candidate materials with both deformability and thermoelectric potential were successfully identified. Notably, density functional theory (DFT) calculations have confirmed that $NbSe_2Br_2$ exhibits a maximum ZT value of 1.35 at 1000 K, which is currently the highest record among known flexible thermoelectric inorganic materials. At 300K, the PF reaches 8.1 $\mu Wcm^{-1}K^{-2}$, second only to the best known flexible TE materials $SnSe_{1.95}Br_{0.05}$[6]. This study promises new discovered inorganic semiconductors toward flexible TE and is expected to accelerate the development and application of flexible TE devices.

## METHODS
### Machine learning

A dataset comprising 483 ZT values across varying temperatures for 34 vdW layered chalcogenides was curated from published literature using the Starrydata2 web system[21] (see **Table S1** and **Figure S1** in **Supplementary Information**). The dataset included 24 initial features, categorized into 8 elemental, 12 structural, and 1 temperature-related attributes (see **Table S2** in **Supplementary Information**), sourced from the JARVIS database[22] and Mendeleev packages. Pearson correlation analysis identified high collinearity among lattice constants (see **Figure S2** in **Supplementary Information**), resulting in the removal of redundant features and retaining 21 relevant descriptors. This processing yielded an input matrix of 10,143 data points (483 samples × 21 features). Six ML algorithms including Decision Tree[23], Random Forest[24], Gradient Boosting Regressor[25], AdaBoost[26], XGBoost[27], and LightGBM[28] were implemented using Scikit-learn[29]. Model performance was evaluated through 10-fold cross-validation, employing root mean square error (RMSE) and coefficient of determination ($R^2$) as metrics to determine the optimal model for predicting ZT values. More details can be found in the **Supplementary Information**.

### First-principles calculations

All the DFT calculations were performed using the Vienna Ab initio Simulation Package (VASP)[30,31]. Projector augmented wave (PAW)[32] method was used, and Perdew–Burke–Ernzerhof (PBE) generalized gradient approximation (GGA)[33] exchange correlation functional was chosen. A plane-wave cutoff of 600 eV and k grids of 5 × 5 × 4 were applied for unit cell calculations. Atomic relaxation was performed until the force on each atom is smaller than 0.0001 eV Å$^{-1}$, and the total energy change was less than $10^{-8}$ eV. Taking into consideration the inclusion of vdW forces, the results were calculated at the optB88 level[34].

The phonon dispersions are computed using the finite displacement method[35] with the Phonopy code[36] and the 2 × 2 × 2 supercells. The temperature-dependent effective potential (TDEP) method is used to extract anharmonic force constants[37,38]. This is done to provide a stable well constrained interatomic force constants (IFCs) for the complex crystal structure of these compounds. The TDEP calculation is based on Born–Oppenheimer molecular dynamics with the PAW method at 300 K with Nose thermostat temperature control[39]. A simulation time of 1000 step, with a time step of 2 fs for each of the ten 2 × 2 × 2 canonical configuration supercell structures generated by TDEP codes. The convergence of the third-order cutoff and the q grids were tested as shown in **Supplementary Information Figure S3**.

Because the PBE functional generally underestimates the bandgap of a semiconductor, we further adopted the hybrid functional HSE06[40] to correct the bandgap. AMSET[41] code was used to solve Boltzmann transport equation and calculate the values of thermoelectric coefficients. Band gaps are corrected using a scissor operation to match those calculated by the HSE06 functional. Piezoelectric constants, and static and high-frequency dielectric constants were computed using DFPT based on the method developed and by Baroni and Resta[42] and adapted to the PAW formalism by Gajdoš et al.[43]. The electron relaxation time $\tau_e$ is calculated by including the fully anisotropic acoustic deformation potential (ADP) scattering, polar optical phonon (POP) scattering, and ionized impurity (IMP) scattering.

**RESULTS AND DISCUSSION**
**High-Throughput Screening of vdW Layered Chalcogenides with High ZT Values and Deformability**

Starting with the JARVIS[22] database, which contains over 70,000 materials, including 3,440 vdW layered materials, we initially screened out 1,179 types of chalcogenides (**Figure 1a**). Machine learning model with high accuracy was developed to efficiently evaluate the thermoelectric performance of the identified candidate materials. Then, based on the mechanical stability criteria, which requires the positive definite elastic constants matrices[44], and the value of the elastic constants of the material should not be too small to avoid numerical errors[20], only 182 kinds of materials were further selected. To identify candidate materials with high deformability, we predicted the deformability of these candidates using our previously developed deformability factor[20], as follows:

$$\theta = \frac{C_{33} \cdot C_{66}}{(\max(C_{11}, C_{22}))^2 \cdot \min(C_{44}, C_{55})} \quad (1)$$

where $C_{ii}$ ($i = 1, 2, 3$) is elastic constant. Materials with deformability factor that is larger than 0.02 were identified as candidates with high deformability. Combined with the criterion that ZTmax should be greater than 1, the candidate materials for flexible TE were finally identified.

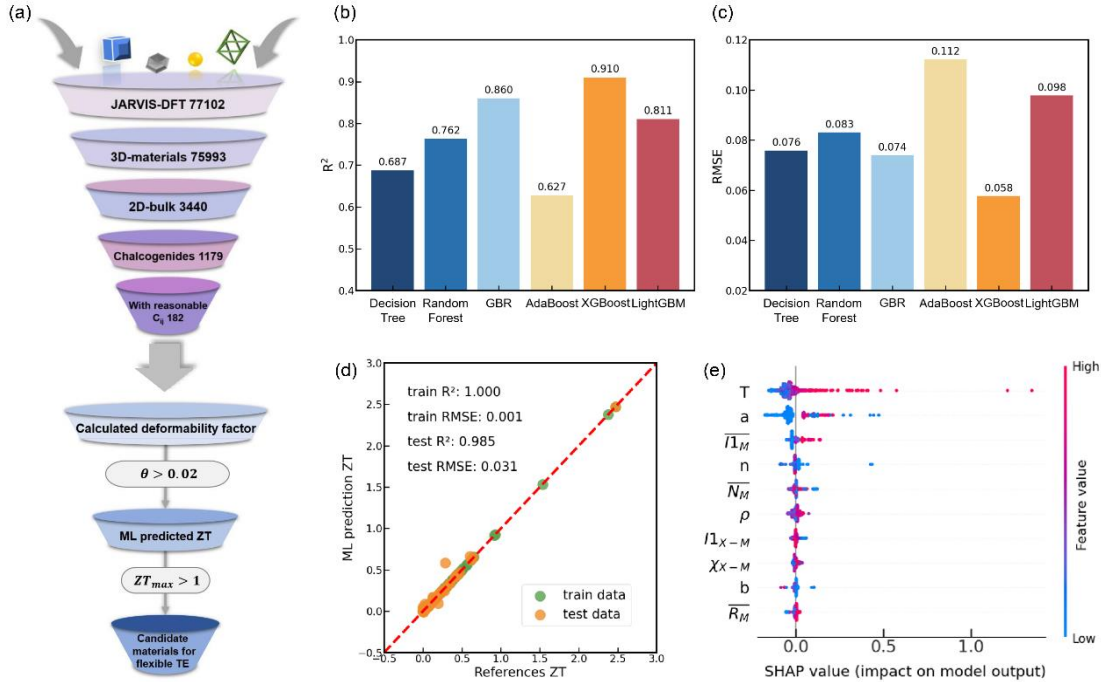

**Figure 1**. (a) Work flow of high-throughput screening for candidate materials with both high deformability and TE performance. The 10-fold cross-validation (b) $R^2$ and (c) RMSE scores of the sequential feature selector results for the six selected ML models, (d) comparison between ZT values obtained from references and ML prediction for train and test datasets, (e) the relatively important value of the top 10 features and the monotonicity between the features and ZT obtained through SHAP.

The machine learning model was developed based on a dataset comprising 483 ZT values measured across various temperatures for 34 van der Waals (vdW) layered chalcogenides (more details see the **METHODS** section). The dataset was divided into 75% training and 25% testing sets. As shown in **Figure 1b-c**, the XGBoost algorithm exhibits the best fitting performance among the six ML algorithms. The $R^2$ and RMSE values for the train set and test set were 1, 0.001 and 0.985, 0.031, respectively (**Figure 1d**). These results demonstrated the validity of the ML model in predicting the ZT values at various temperatures, and confirmed its robust reliability. Additionally, the SHapley Additive exPlanations (SHAP)[45] technique was used to obtain the relatively important value of the features and identify the monotonicity between the features and ZT. It was found from **Figure 1e** that the temperature $T$, lattice constant $a$, first ionization energy of non-chalcogenide elements $\overline{I1_M}$, and number of atoms in the primitive cell $n$ are the four most important features.

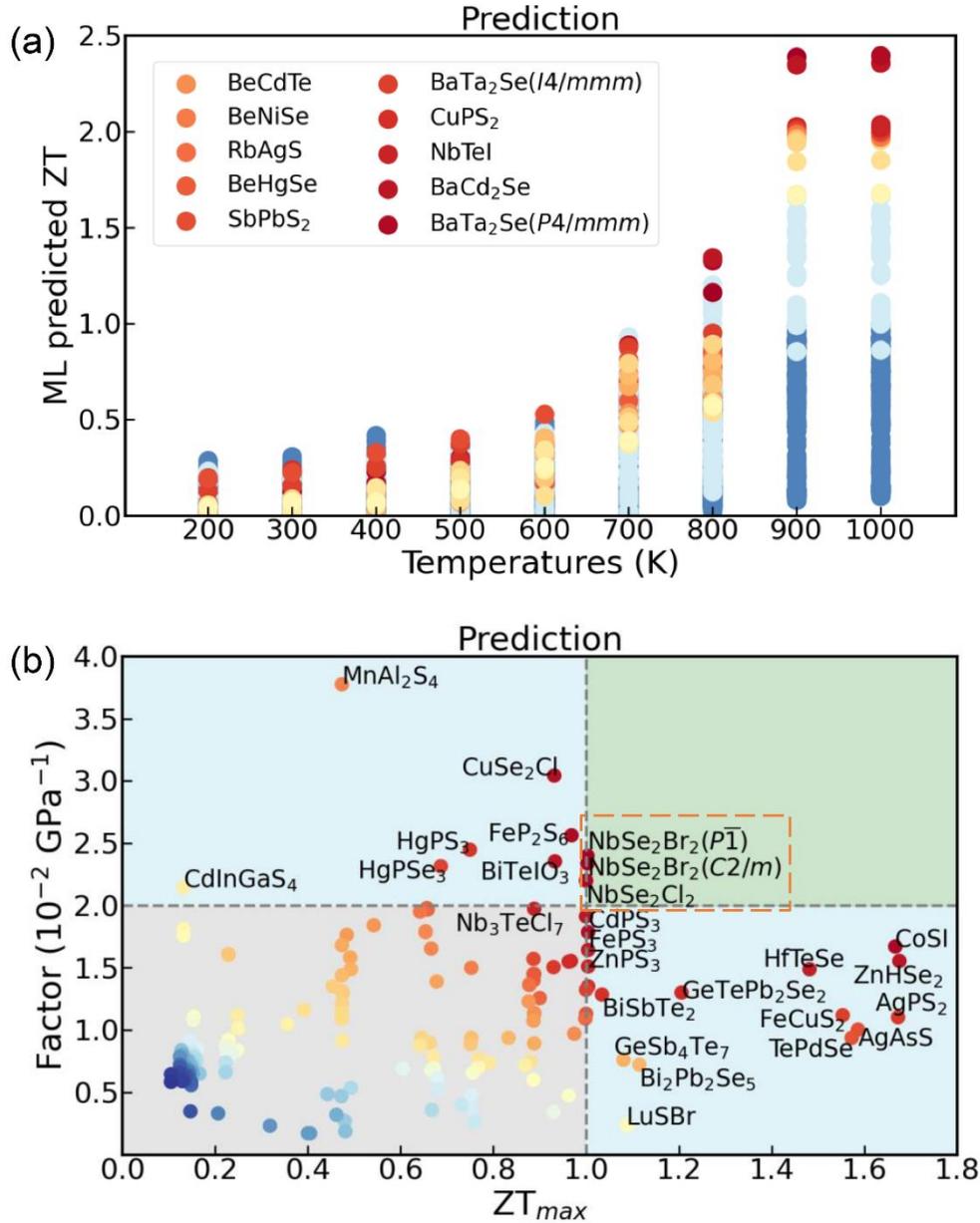

**Figure 2**. (a) The ZT of the unexplored vdW layered chalcogenides predicted by the ML model, and (b) the deformability factors as well as $ZT_{max}$ of the screened candidates.

**Figure 2** shows the ZT of the unexplored vdW layered chalcogenides predicted by the ML model, and the deformability factors as well as $ZT_{max}$ of the screened candidates. As can be clearly seen from **Figure 2a**, the ZT values increase with increasing temperature, and there are many kinds of materials exhibit high ZT values at high temperatures, which may be promising TE candidates. The top 10 candidates are marked, that are all ternary chalcogenides, showing the thermoelectric potential of this class of compounds and deserving further study. Materials with deformability factors larger than 0.02 and $ZT_{max}$ larger than 1 were screened as candidates for flexible TE materials and plotted as **Figure 2b**. Unfortunately, some materials with high deformability exhibit low $ZT_{max}$, as shown in the upper left corner in **Figure 2b**. They may be good flexible inorganic materials for flexible electronics and wearable devices. Similarly, those materials with high ZTmax show low deformability factors, distributed in the lower right corner. Although they are brittle, they may be good TE materials for TE cooling, TE power generation, etc. Materials with both high deformability factors and TE potential are very rare, showing a large blank area in the upper right corner. There are only some possibilities on the borderline, which may be worth further investigation. (see **Table S3** in **Supplementary Information** for more details). Among them, we highlight the material NbSe$_2$Br$_2$ as the most promising candidate. Specifically, NbSe$_2$Br$_2$ exhibits two phases, both with the same ZT$_{max}$ (1.0026 at 1000 K). However, the triclinic $P\bar{1}$ (No. 2) phase holds a slightly higher deformability factor ($2.4 \times 10^{-2} GPa^{-1}$) compared to the monoclinic $C2/m$ (No. 12) phase ($2.3 \times 10^{-2} GPa^{-1}$).

**Structure of the Identified Candidate NbSe$_2$Br$_2$**

Considering that the triclinic $P\bar{1}$ phase of NbSe$_2$Br$_2$ has a higher deformability and lower formation energy (-0.77231 eV compared to -0.77226 eV for the C2/m phase), we choose the NbSe$_2$Br$_2$ with $P\bar{1}$ space group for further DFT calculation validation (see details in **Methods**). The calculated lattice constants are shown in **Table 1**, which are consistent with the data from JARVIS[22] database. In **Figure 3a**, it is obvious that the NbSe$_2$Br$_2$ crystal has parallel ab-plane sheets bonded by vdW interactions. The single-layer structure is flat, which may be prone to interlayer slip and conducive to flexible deformation. Each niobium ions Nb$^{4+}$ forms a polyhedron with four selenium ions Se$^{2-}$ and four bromide ions Br$^-$. Adjacent polyhedrons are connected by common Br-Br bonds or common faces consisting of four Se$^{2-}$ ions. All the Br$^-$ ions are bonded to Nb$^{4+}$ ions, and all the Se$^{2-}$ ions are dimerized, forming Se$_2^{2-}$ dimers between Nb$^{4+}$ ions. **Figure 3b** shows the ab-plane of triclinic NbSe$_2$Br$_2$, where there are loops consists of 6 Nb$^{4+}$ ions in total. This hollow ring structure may lead to weaker in-plane elastic properties, thus facilitating flexible deformation.

**Table 1**. Lattice constants of NbSe$_2$Br$_2$.

| a (Å) | b (Å) | c (Å) | α (°) | β (°) | γ (°) | Source |
|---|---|---|---|---|---|---|
| 6.72 | 6.79 | 7.43 | 67.65 | 67.88 | 60.38 | JARVIS[22] database |
| 6.7005 | 6.7839 | 7.1361 | 67.6828 | 67.9030 | 60.3637 | This work |

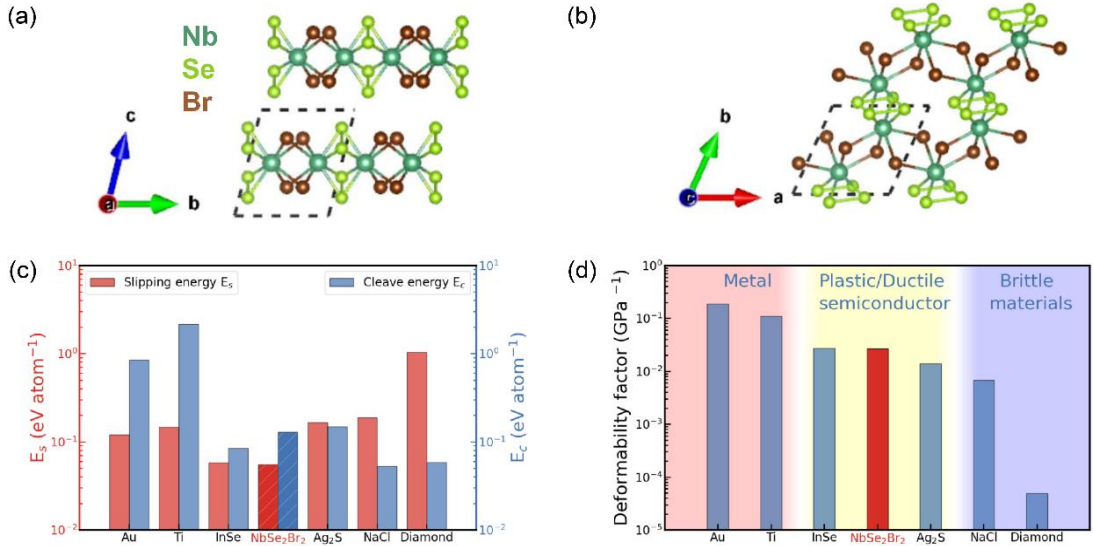

**Figure 3**. Atomic structures along (a) a-axis and (b) c-axis. (c) Slipping energy barrier $E_s$, cleavage energy $E_c$, and (d) deformability factors $\Xi$ along the slipping system (001) <100> for $NbSe_2Br_2$. The data of Au, Ti, InSe, $Ag_2S$, NaCl, and diamond are also included for comparison[13].

**Further Confirmation of the Deformability Properties of $NbSe_2Br_2$**

We calculated the elastic constants of $NbSe_2Br_2$ and compared with typical plastic vdW material InSe as shown in **Table 2**. $NbSe_2Br_2$ has similarly small in-plane elastic constants $C_{11}$ and $C_{22}$ as InSe, which may mean that it is more likely to deform in-plane and less likely to break, thus facilitating flexible deformation. This characteristic may be related to the hollow ring structure in the plane mentioned above. At the same time, similar to InSe, the $C_{33}$ of $NbSe_2Br_2$ is not very small compared with the in-plane elastic constants $C_{11}$ and $C_{22}$, which may mean a strong interlayer interaction, resulting in a highe cleavage energy similar to InSe. Moreover, the $C_{44}$ and $C_{55}$ is very small compared to $C_{66}$, which is also similar with InSe, and means a weak shear modulus, so interlayer slip is easy to occur, which is conducive to flexible deformation. This may due to the flat single-layer structure shown in **Figure 3a**.

With the calculated elastic constants and our previously developed method[20], we calculated the deformability factor of $NbSe_2Br_2$, which is comparable to that of the typical plastic vdW layerd chalcogenide InSe, showing promising deformability. We further verified with the method in the previous research[13] and compared $NbSe_2Br_2$ with typical mental materials, plastic/ductile semiconductors and brittle materials. As shown in **Figure 3c**, $NbSe_2Br_2$ do shows high cleavage energy and low slipping energy just like InSe. $NbSe_2Br_2$ finally shows deformability factor much higher than typical brittle materials like diamond, and comparable with typical mental materials with excellent plasticity like Au and Ti, validating the flexibility again (**Figure 3d**).

**Table 2**. Mechanical properties of InSe and $NbSe_2Br_2$. (The unit of elastic constants and deformability factor is GPa and $10^2 \times GPa^{-1}$, respectively)

| $C_{11}$ | $C_{22}$ | $C_{33}$ | $C_{44}$ | $C_{55}$ | $C_{66}$ | Factor | Source |
|---|---|---|---|---|---|---|---|
| 60.4 | 60.4 | 34.5 | 7.7 | 7.7 | 20.2 | 2.480881 | InSe[20] |
| 66.1 | 82.3 | 28.8 | 6.6 | 7.9 | 37.3 | 2.403022 | JARVIS[22] database |
| 65.99 | 82.17 | 28.75 | 6.58 | 7.82 | 37.06 | 2.397929 | This work |

**Electronic and phonon properties**

To check the semiconductor characteristics of this structure, **Figures 4a-b** show the calculated electronic band structure of NbSe$_2$Br$_2$. The calculated indirect band gaps using the PBE and HSE06 functionals are 0.87 eV and 1.35 eV, respectively. The density of states (DOS) reveal that the Nb-d states contribute more to the valence band region, while the Se-p states dominate in the conduction band region. Additionally, the flat nature of the valence band and the dispersive conduction band suggest promising thermoelectric properties, which will be discussed in the following section.

The phonon spectrum calculation results in **Figure 4c** confirm the dynamic stability of this structure. It can be seen that acoustic phonon branches TA modes are mainly in the low energy range of 0-6 meV, while the LA modes hold the higher energy reaching around 10 meV and entangle with optical phonons in the energy range of 6-10 meV along most of the high symmetry paths except Γ-Z direction. There is no acoustic-optical phonon gap along these high symmetry paths, suggesting a strong coupling effect between the acoustic and low-lying optical phonons, which could provide massive phonon-phonon scattering channels for scattering processes[46]. Along Γ-Z direction, the three acoustic phonon branches are all in the very low energy range of 0-5 meV and do not cross any optical phonons, resulting in an acoustic-optical phonons gap there. These features suggest that there may be extremely low lattice thermal conductivity $\kappa_l$ along Γ-Z direction, which is very common in vdW layered materials and will be discussed below.

From the phonon projected density of states (PDOS) plotted in **Figure 4d**, the acoustic phonons are mainly contributed by the chalcogen element Se, and the low-lying optic phonons lower than 25 meV are mainly contributed by the halogen element Br. There are many optical phonons distributed in the range of 10-22 meV, most of which are flat. Two optical-optical phonons gaps exisit in the range of 21-28 meV and 33-40 meV, respectively. Optical phonons in the high energy range above the gaps are even flatter and may have very low group velocities, thus, may have little significant contribution to the $\kappa_l$, which will also be discussed later. Above the optical-optical phonons gap, the optical phonons in the energy range of 25-35 meV are mainly contributed by the transition metal element Nb, while the highest optical phonons in the energy range of around 40 meV are also mainly contributed by Se, as shown in **Figure 4d**.

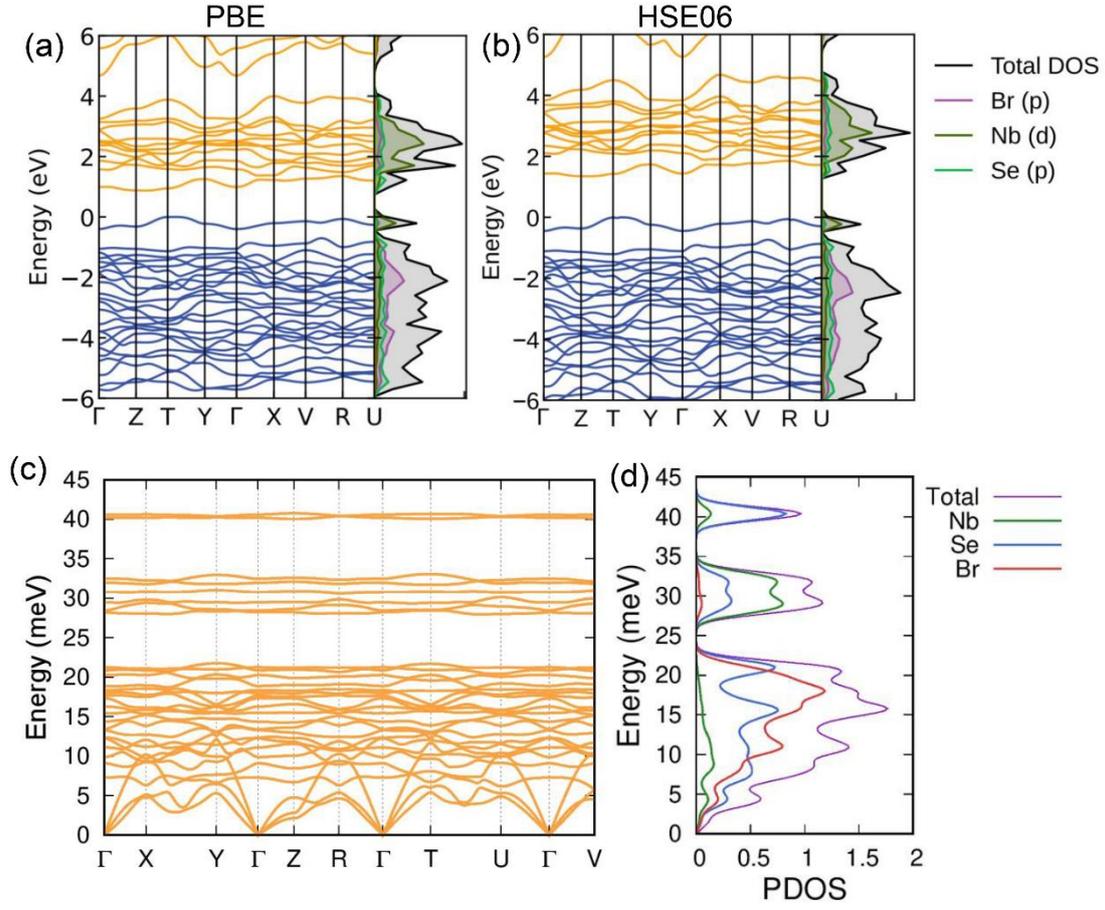

**Figure 4**. The calculated electronic band structures and density of states using (a) PBE and (b) HSE06 functional. (c) Phonon dispersion and (d) phonon projected density of states (PDOS) of $NbSe_2Br_2$.

**Lattice thermal conductivity of $NbSe_2Br_2$**

**Figure 5a** shows the calculated temperature dependent $\kappa_l$ which exhibits obvious anisotropy. $NbSe_2Br_2$ shows very low $\kappa_l$ especially along the *z*-axis (0.53 W m$^{-1}$K$^{-1}$ at 300 K). These ultra-low $\kappa_l$ is conducive to high TE performance. As shown in **Figure 5b**, the energy dependence of the cumulative $\kappa_l$ along the *x*-axis and *y*-axis increase rapidly until 10 meV and then be almost flat. The steep growth under 10 meV indicates the large contribution of acoustic phonon modes shown in **Figure 5c** to $\kappa_l$, while the obvious slowing down trend indicates that the optical modes contribute little to $\kappa_l$ as mentioned above. It's interesting to find that the slope of the cumulative $\kappa_l$ curve along the *x*-axis is as high as that along the *y*-axis under 5 meV and higher in the energy range of 5-10 meV, indicating that the anisotropy between these two axises is caused by phonons in this energy range. Along *z*-axis, the cumulative $\kappa_l$ increases rapidly only under 5 meV, and then flattening out. The slope of the growth curve is also much smaller than those along the *x*-axis and *y*-axis, indicating that the thermal conduction along this axis is weaker than those along the *x*-axis and *y*-axis, which also leads to a significantly lower $\kappa_l$.

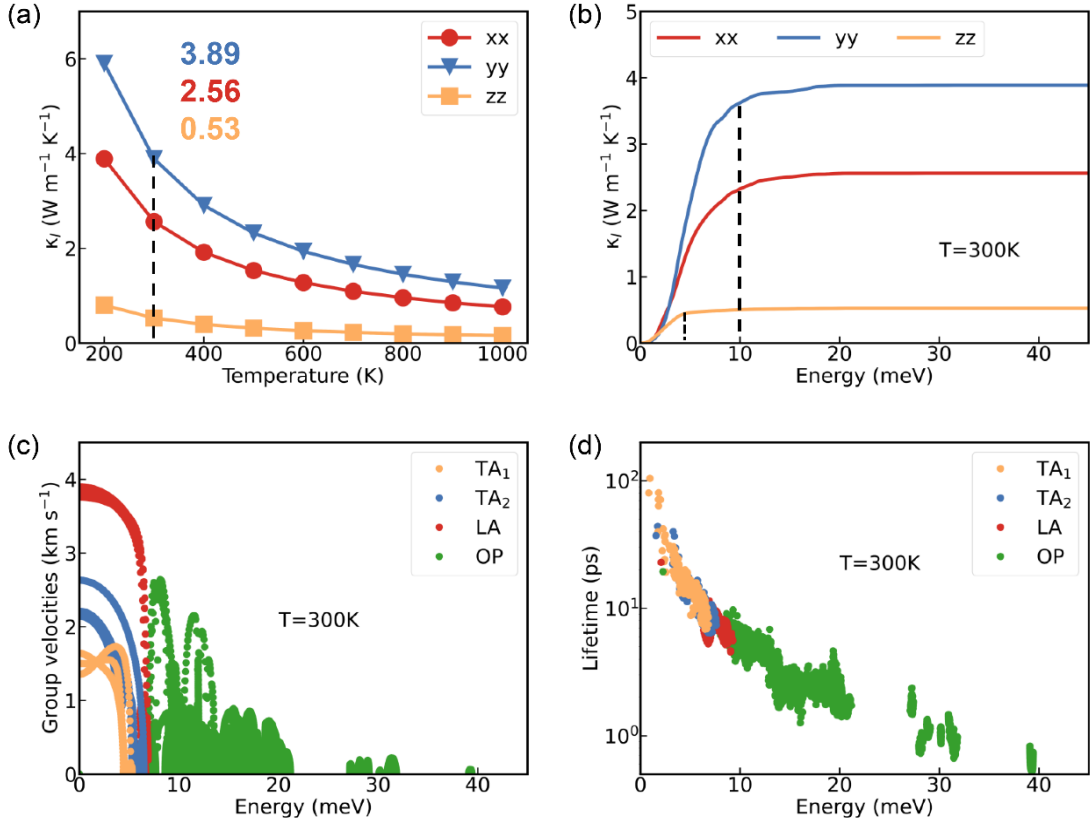

**Figure 5**. (a) Temperature dependent, (b) energy dependent $\kappa_l$, (c) group velocity, and (d) phonon lifetime at 300K of NbSe$_2$Br$_2$.

To further investigate the phonon and thermal conduction properties, we calculated the mode-resolved group velocity and phonon lifetime at 300 K. As shown in **Figure 5c**, only LA modes shows high group velocity, while the group velocity of TA modes and low-lying optical modes are less than 3 km s$^{-1}$. Meanwhile, the group velocity of optical modes in the energy range of over 15 meV do not exceed 1 km s$^{-1}$. It was found from **Figure 5d** that the phonon lifetime is very short, most of which is lower than 100 ps. As the energy increases, the phonon lifetime decreases significantly. Compared with acoustic phonons, the lifetime of optical phonons in the energy range of above 15 meV is more than an order of magnitude shorter. The obviously small group velocity as well as the significantly short phonon lifetime of high-energy optical modes make their contribution to $\kappa_l$ very limited, as mentioned above in the discussion on **Figure 5d**. Thus, it was the small group velocity and short phonon lifetime that mainly contributes to the ultra-low $\kappa_l$ of NbSe$_2$Br$_2$.

**Figure of merit ZT of NbSe$_2$Br$_2$**

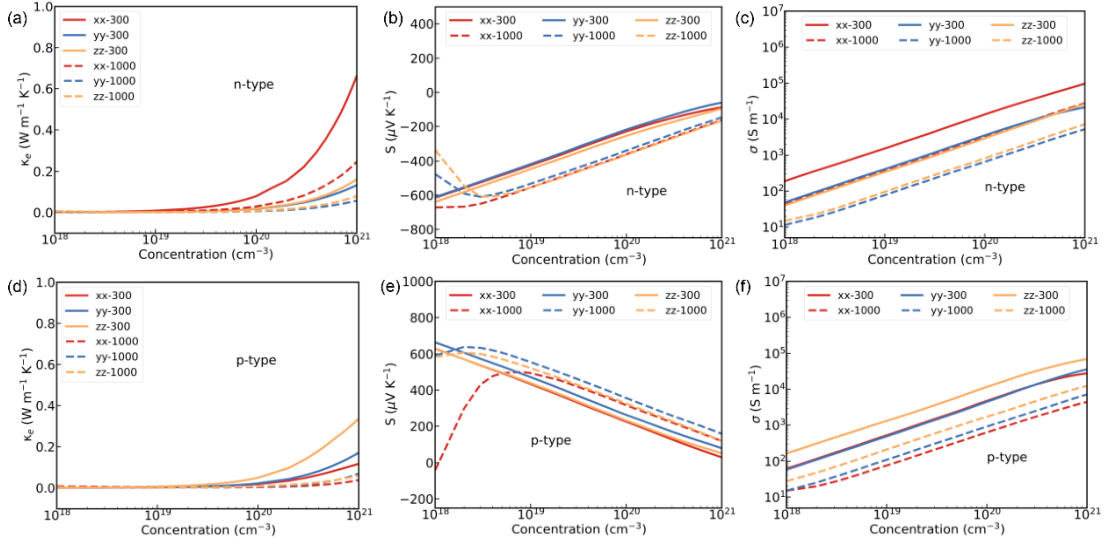

**Figure 6**. The calculated (a, d) electronic thermal conductivity ($\kappa_e$), (b, e) seebeck coefficient (S), and (c, f) electrical conductivity($\sigma$) of NbSe$_2$Br$_2$ along *x*-, *y*-, and *z*-axis at 300 K and 1000 K for n-type and p-type concentration from 1E18 to 1E21 cm$^{-3}$.

We use AMSET[41] code to solve Boltzmann transport equation and calculate the values of thermoelectric coefficients. The estimated electronic thermal conductivity ($\kappa_e$), seebeck coefficient (S), and electrical conductivity($\sigma$) of NbSe$_2$Br$_2$ at 300 K and 1000 K for n-type and p-type concentration from 1E18 to 1E21 cm$^{-3}$ are shown in **Figure 6**. Except for the high $\kappa_e$ along *x*-axis at 300 K for n-type, the $\kappa_e$ is very low in other cases, but cannot be ignored compared with the $\kappa_l$ at high doping concentrations. Above 1E19 doping concentration, the S of n-type increases with the increase of doping concentration, while that of p-type decreases with the increase of doping concentration. The S is larger at high temperature 1000 K than at low temperature 300 K for both n-type and p-type, and there is no obvious anisotropy between different directions as shown in **Figure 6b and 6e**. The $\sigma$ is highest along *x*-axis at 300K for n-type, while it is highest along *z*-axis at 300K for p-type. For both n- and p-type, the $\sigma$ increases with the doping concentration.

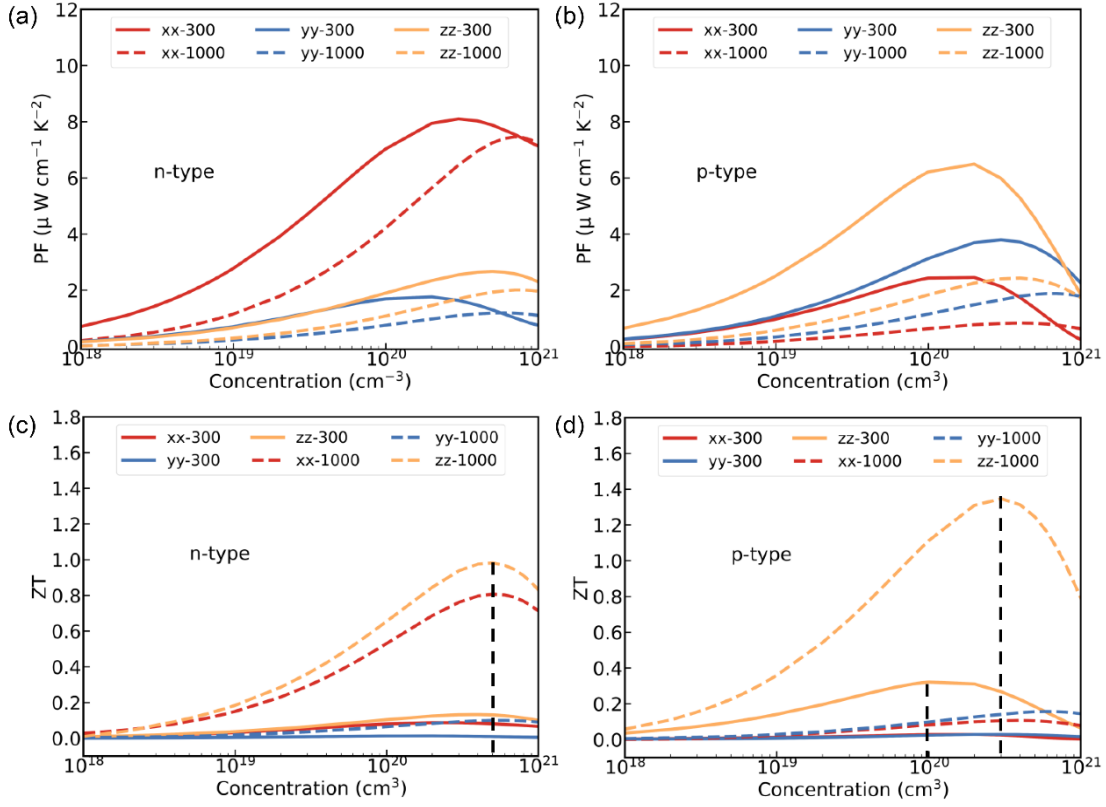

**Figure 7**. The calculated (a, b) PF, and (c, d) ZT of NbSe$_2$Br$_2$ along *x*-, *y*-, and *z*-axis at 300 K and 1000 K for n-type and p-type concentration from 1E18 to 1E21 cm$^{-3}$.

**Figure 7a-b** show the PF of which electrons are higher than holes. The maximum PF observed is 8.1 μ W cm$^{-1}$ K$^{-2}$ along *x*-axis at 300K in n-type doping concentration 3E20, surpassing most organic and inorganic flexible thermoelectric materials and second only to the best known flexible TE materials SnSe$_{1.95}$Br$_{0.05}$[6], as shown in **Figure 8a**. The calculated ZT values are shown in **Figure 7c-d**. At 1000 K, the maximum ZT is reached 1.35 at a p-type doping concentration of 3E20 along *z*-axis. And the maximum ZT observed at 300K is 0.32 at a p-type doping concentration of 1E20 also along *z*-axis. In the n-type doping concentration 5E20, the ZT$_{max}$ reaches at a high temperature of 1000K along both *x*- and *z*-axis, and the values are both close to 1. It is worth noting that in the field of flexible TE, the ZT$_{max}$ value 1.35 of NbSe$_2$Br$_2$ is currently a recorded high value compared with other flexible inorganic TE materials discovered so far, as shown in **Figure 8b**.

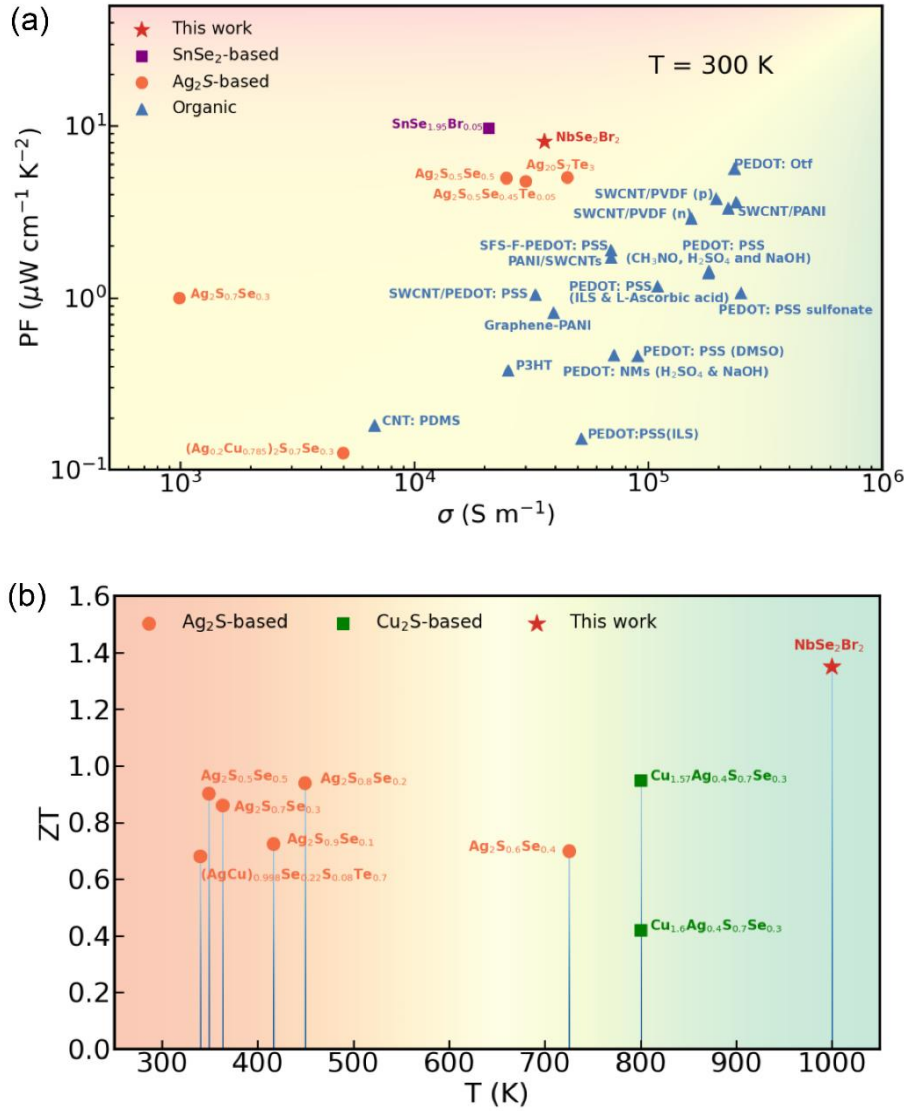

**Figure 8**. (a) PF at 300 K as a function of $\sigma$ for NbSe$_2$Br$_2$, flexible inorganic vdW layered crystal SnSe$_2$-based TE materials, flexible inorganic Ag$_2$S-based TE materials, and typical flexible organic TE materials. (Data are taken from reference[6]). (b) ZT as a function of temperatures for NbSe$_2$Br$_2$, flexible inorganic Ag$_2$S-based and Cu$_2$S-based TE materials. (Data are taken from references[10,47]).

**CONCLUSION**

In summary, we high-throughput screened over 1000 vdW layered chalcogenides from material databases and predicted their deformability with our previously developed deformability factor. A highly accurate ML model with a $R^2$ of 0.985 and RMSE of 0.031 were successfully developed to predict the temperature dependent ZT values, with which several candidate materials for flexible TE application with both high deformability and TE performance were discovered. More importantly, the top-ranked compound NbSe$_2$Br$_2$ was verified by DFT calculations to exhibit a maximum ZT value of 1.35 at 1000 K, which is currently a recorded high value compared with other flexible inorganic TE materials discovered so far. What's more, the PF value of

8.1 μWcm$^{-1}$K$^{-2}$ at 300K also surpassed most organic and inorganic flexible thermoelectric materials. Its high deformability mainly comes from the small slipping energy that allows interlayer slip and the small in-plane modulus that allows deformation before failure. The high ZT$_{max}$ is mainly attributed to the extremely low thermal conductivity and the high Seebeck coefficient along the out-of-plane direction at high temperature. The high power factor at room temperature is mainly contributed by the high conductivity in the in-plane direction. This work provides not only promising vdW layered chalcogenides for further flexible TE applications but also an efficient method which is applicable to the discovery of various types of functional materials.

**ACKNOWLEDGMENTS**
Q. R. was supported by Graduate Technological Innovation Project of Beijing Institute of Technology (Grant No. 2023YCXZ002). Yingzhuo Lun was supported by National Natural Science Foundation of China (Grant No. 12402183), Beijing Natural Science Foundation (Grant No. 1244057).